\documentstyle[aps]{revtex} 
\begin{document}

\renewcommand{\thesection}{\Roman{section}}

\centerline{\Large{\bf{Synchronization in Coupled Sine Circle Maps}}}
\vspace{0.4in}
\centerline{Nandini Chatterjee and Neelima Gupte}
\vspace{0.2in}
\centerline{\it{Department of Physics, University of Pune, Pune 411 007, India}}\vspace{0.2in}
\begin{abstract}
We study the spatially synchronized and temporally periodic solutions of a 
1-d lattice of coupled sine circle maps.
We carry out an analytic stability analysis of this spatially synchronized
and temporally periodic case and obtain the stability
matrix in a neat block diagonal form. We find spatially synchronized behaviour
over a substantial range of parameter space. We have also extended
the analysis
to higher spatial periods with similar results. Numerical simulations
for various temporal periods of the synchronized solution, reveal that
the entire structure of the Arnold tongues and the devil's staircase
seen in the case of the single circle map can also be observed for the
synchronized coupled sine circle map lattice. Our formalism should
be useful in the study of spatially periodic behaviour in other 
coupled map lattices.
\end{abstract}
\vspace{10pt}
PACS number(s) : 05.45. +b

\newpage 

The study of complex dynamical behaviour in extended systems is  currently of 
interest 
in a wide variety of contexts.  
The modelling and characterization of spatiotemporal behaviour in such systems
 \cite{kaneko 93,crutch 87} can provide insights into the complex behaviour  
found in diverse systems like  oscillator arrays
\cite{choi 94,op 86,wa 84,yam 84,bus 85,kaneko 89}, coupled Josephson junction
 arrays 
\cite{kwies 89}, reaction diffusion systems \cite{kap 93,davi 92}, charge
 density waves \cite{levy 92}, biological systems \cite{nag 72,free 85} and
  turbulence 
in fluids \cite{shree 85}. The spatially extended nature of the system permits
 the appearance of complex spatio-temporal behaviour  
such as spatially periodic, quasi-periodic or chaotic behaviour with the 
concurrent appearance of temporally periodic, quasi-periodic or chaotic
 behaviour.  
An understanding of the rich variety of modes which can be excited due to the
 interplay between 
spatial and temporal behaviour may provide a  
clue to phenomena like   
pattern formation in natural systems \cite{kaneko 93,bar 91,swine 91} and
 turbulence. 
 
Coupled Map Lattice (CML) models have recently attracted much attention in
 the study of spatio-temporal chaos and pattern formation as models of 
  spatially extended systems \cite{kaneko 93,crutch 87}. A CML is a dynamical 
  system with discrete time, discrete space and continuous states. It
    consists  of dynamical elements on a lattice which are suitably coupled.
     Such systems  have been used succesfully to model real life
phenomena like spiral waves \cite{bar 91} and spatio-temporal intermittency
 \cite{ch 88}.    
Though CML models are idealized systems, they are sufficiently complex to be
 capable of capturing  the essential features of the dynamics of the system,
  and at the same time  have the advantage of being mathematically tractable 
  and computationally  efficient.

Due to the large number of degrees of freedom in such spatially extended 
systems, a variety of spatio-temporal phenomena, like synchronization, 
intermittency, and spatio-temporal chaos are observed. 
One of the most important and interesting modes which can arise in such 
systems is the mode corresponding to synchronised behaviour, i.e. behaviour in 
which
evolution at each spatial location is identical with that at every other
spatial location at any arbitrary instant of time. (SeeFig.1a) Such 
synchronized behaviour has been observed in a number of spatially extended 
systems \cite{gade 95}, like  coupled oscillator arrays  
\cite{wa 84,kwies 89,heagy 94}, coupled pendulums \cite {abraham 93}, 
electronic oscillator circuits \cite {ash 90} and in pattern formation 
\cite {kk 89}. As 
coupled sine circle map systems \cite{crutch 87} can constitute models which 
capture many  
of the essential features of the behaviour of such systems,   
we address the problem of synchronized or spatially homogeneous solutions in 
a 1-dimensional array of coupled sine circle maps. 
The temporal evolution of these states may be periodic, quasi-periodic or 
chaotic in nature. The present paper concentrates on synchronized states 
that are temporally periodic in behaviour.   
 
The single circle map \cite{bak 20,bak 84},is represented by  
\begin{equation} 
\theta_{t+1}= f(\theta_{t}) = \theta_{t} + \Omega 
-{K \over(2 \pi)} \sin(2 \pi\theta_{t}) \hskip 0.2in   mod \hspace{0.05 in}1 
\end{equation} 
where $\theta_{t}$ is an angle, at time $t$ which lies between 0 and 1, 
$K$ is the nonlinearity parameter and $\Omega$ is the period of the system 
for $K=0$.
This is one of the simplest representations of physical phenomena involving 
periodic motion. This simple dynamical system which exhibits a tendency to 
mode lock, as the parameter $K$ is increased, is particularly suitable for 
the description of resonances between periodic motion \cite{bak 20}. 
An extensive study by Bak {\it et al} \cite{bak 84}, shows the presence of 
the Arnold tongues in the $\Omega-K$ space and complete mode-locking, 
namely the Devil's staircase at $K = 1$. 
Since the single sine circle map exhibits this tendency to mode-lock, it
is interesting to study whether an array of such sine circle
maps on a lattice suitably coupled, would demonstrate such a 
mode-locking spatially. Synchronization is the simplest example of such
a phenomenon and we focus
our attention on the synchronized states which are temporally periodic in an 
array of such coupled
circle maps.

The specific model under study, is a 1-dimensional coupled map lattice of 
sine-circle maps with nearest neighbour diffusive symmetric normalized 
coupling (also called 
future coupled laplacian coupling) and periodic boundary conditions, and is 
given by 
 
\begin{eqnarray} 
 \theta_{t+1}(i)&  = & (1 - \epsilon)f(\theta_{t}(i))
 + {\epsilon \over 2}f(\theta_{t}(i+1)) + {\epsilon \over 2} f(\theta_{t}(i-1)) 
 \hspace{0.05 in} mod \hspace{0.01 in} 1\nonumber \\ 
   & = &  (1-\epsilon) \Big( \theta_{t}(i) + \Omega - 
{K \over(2 \pi)} \sin(2 \pi\theta_{t}(i)) \Big) \nonumber \\ 
 & + &{\epsilon \over2}   
\Big\{\theta_{t}(i+1) 
  +  \Omega -{K \over(2 \pi)} \sin(2 \pi\theta_{t}(i+1)) \nonumber \\ 
  &+ & 
\theta_{t} 
(i-1) + \Omega   
 - {K \over(2 \pi)} \sin(2 \pi\theta_{t}(i-1)) \Big\} \hspace{0.2 in} mod \hspace{0.05 in} 1
\end{eqnarray} 
where $\theta_{t}(i)$ is the angular variable associated with the $i$th site, 
at time $t$. \\ 
The parameters $ \Omega $ and $ K $ are
taken to be uniform at each site and defined as in Eq.2 for the single circle 
map, \\
and $\epsilon$ which lies between 0 and 1 is the strength of the coupling 
parameter. 
 
We investigate in detail the phenomena of spatial synchronization in a system 
of such coupled sine circle maps. 
We identify the spatially homogeneous and temporally periodic, 
quasi-periodic and chaotic modes of the system.  
 We carry out a linear stability analysis to analyse the stability properties 
 of the spatially synchronized, temporally periodic solutions. The 
 independent variables
of the problem 
are identified and  
the analysis is cast in terms of these independent variables. This leads to a 
neat and simple block diagonal form for the stability matrix. This permits us 
to identify the regions in parameter space where synchronized solutions  
of different temporal periods are stable.The limits of the mode-locked 
interval for the temporal period one case can be explicitly evaluated 
due to the fact that the stability matrix has a block diagonal, block
circulant structure, for this case.
This analysis is also extended to higher spatial periods. Our  
method is quite general and should prove to be useful in the study of
spatially periodic behaviour in other coupled map
lattice models as well.   
 
The paper is divided into five sections. Section I discusses the coupled 
shift map, which is the linear version  of the coupled circle map, 
namely $K = 0$ in Eq .2, and is thus the simplest case. 
We identify the independent variables for this system and recast the
equation of evolution in terms of these variables. We carry out a systematic 
linear
stability analysis, and obtain a neat form for the
stability matrix. In section II, we extend the formalism to the system
of coupled sine circle maps, taking into account the nonlinear sine term as
defined in Eq. 2 and cast the problem 
in terms of the new independent variables of this system. The linear 
stability analysis in terms of these variables, gives a neat block diagonal 
form for the stability matrix. 
This is followed by an explicit calculation of the eigenvalues 
for the synchronized solution and the evaluation of the limits of the 
mode-locked interval for the fixed point case. 
 Section III presents the  
generalization to higher spatial periods.  
In section IV, the numerical simulations for the higher temporal periods, 
of the
synchronized solution, are discussed. We extend the algorithm developed by
Bak {\it et al} \cite{bak 84}, for the single circle map, to the synchronized
solution of a system of coupled sine
circle maps.Section V concludes with results and a short discussion. 

\section{\bf Analysis for the shift map lattice}
We begin by carrying out the analysis for the simplest case of Eq. 2, 
namely for $K = 0$, which is the case
of coupled shift maps.

The single shift map i.e., the linear case of the single circle
map, namely $K = 0$ in Eq.1 is given by

\begin{equation}
\theta_{t+1} = \theta_{t} + \Omega  \hspace{0.2 in} mod \hspace{0.05 in}1
\end{equation}
 This has periodic orbits for rational values of $\Omega$ i. e. $\Omega = 
 { P \over
 Q}$
 where $P$ and $Q$ are any integers, and quasi-periodic orbits for
 irrational values of $\Omega$.

 Eq. 2, with $K = 0$, for the coupled shift map, for a lattice of $N$ sites, 
 retaining periodic boundary conditions, reduces to
 \begin{equation}
 \theta_{t+1}(i) = (1- \epsilon)\Big( \theta_{t}(i) + \Omega \Big) + 
 {\epsilon \over 2} \Big( \theta_{t}
 (i + 1) + \theta_{t}(i-1) \Big) + 2 \Omega 
 \hspace{0.2 in} mod \hspace{0.05 in} 1
 \end{equation}

 For the synchronized solution, since the value of the variable at all sites 
 is
 the same, we note that differences of
 nearest neighbour variable site values is equal to zero for all neighbours.
  Writing an equation for such a difference for the
  first pair of sites of an array of $N$ sites, for the coupled shift map,
we obtain,
  \begin{eqnarray}
  \theta_{t+1}(1) - \theta_{t+1}(2) & = & (1 - \epsilon) \Big( \theta_{t}(1)
   - \theta_{t}(2) \Big) \nonumber \\
   & + & {\epsilon \over 2} \Big( \theta_{t}(2) - 
   \theta_{t}(3) + \theta_{t}(1) - \theta_{t}(N) \Big)
   \end{eqnarray}
   It is clear that Eq.5 can be expressed completely in terms of the differences
  defined by
\begin{equation}
   a_{t}(i)=  \theta_{t}(i) - \theta_{t}(i+1) 
    \end{equation}

The evolution equation of the differences is given by, 
 \begin{equation}
    a_{t+1}(i) = (1 - \epsilon)(a_{t}(i)) + {\epsilon \over 2} \Big( a _{t}(i+1)
       + a_{t}(i-1) \Big) \hspace{0.2 in} mod\hspace{0.05 in} 1
	\end{equation}
   Thus, for the synchronized solution, of the coupled shift map, it is 
   sufficient to define just $N$ differences, to obtain
   the $N$ equations of evolution in terms of independent variables.

Substituting $a_{t}(i) = 0$, in Eq.7, gives $a_{t+1}(i)= 0$, and
hence shows that $a_{t}(i) = 0$, is a spatial fixed point solution.

  Expanding up to the linear term about this solution leads to
  a stability matrix $J^{shift}_{t}$, given by

\begin{equation}
   J^{shift}_{t} = \pmatrix{ 
(1- \epsilon)& {\epsilon \over 2} & 0 &0 & \cdots &  0 & 0& {\epsilon \over
2} \cr
{\epsilon \over 2} & (1- \epsilon)& {\epsilon \over 2} & 0 & 0 &\cdots & 0 &
 0\cr
0 &  {\epsilon \over 2} & (1- \epsilon)&  {\epsilon \over 2} & 0 & \cdots & 0
& 0\cr
   \vdots&\vdots&\vdots&\vdots&\vdots&\vdots&\vdots&\vdots \cr
   {\epsilon \over 2} & 0 & 0&\cdots &  0 & 0& {\epsilon \over 2} &
  (1- \epsilon)\cr}
\end{equation}
		     This is an $N \times N$ matrix, which is also
 circulant and whose eigen values maybe explicitly obtained
 analytically.
      
      The eigenvalues of $J^{shift}_{t}$ are given by \cite{gad 93,davis 79}
      \begin{eqnarray}
      \lambda_{r} & = & (1- \epsilon) 
       + 
        {\epsilon \over 2} ( \omega_{r}
+ \omega_{r}^{N-1})  
 \end{eqnarray}
  where $\omega_ {r}$ is the $N$th root of unity given by\\
   \begin{equation}
    \omega_{r} = \exp{\Big({2i \pi (r - 1)\over N}\Big)}
     \end{equation}
      On simplifying, this can be written as \\
       \begin{eqnarray}
       \lambda_{r} = (1- \epsilon)
 + \epsilon  \cos ({2 \pi (r - 1) \over N}) 
     && \hspace{0.04 in}r = 1,2, \ldots, N
\end{eqnarray} 

The largest eigenvalue of Eq. 11 namely $\lambda^{largest} \leq 1$ defines the 
stability condition for temporally stable and spatially synchronised periodic 
orbits of the coupled shift map. 
The largest eigen value is +1, indicating that the coupled shift map is 
marginally stable. We find that as long as the differences between
nearest neighbour sites tend to zero, we obtain synchronized
solutions, irrespective of the corresponding temporal period.
Since, the largest eigen value does not depend on $\epsilon$, we
conclude that, for the synchronized solution, there is no dependence of the 
coupling parameter 
on the temporal
behaviour. 
Thus for the synchronized solution, the coupled shift map continues to have 
periodic orbits for rational values of
   $ \Omega$, and quasi-periodic orbits for $ \Omega$ irrational, 
  similar to the single shift map and there is no enlargement in the 
  phase space due to the coupling.

\section{\bf Linear stability analysis for the coupled circle map lattice}
We now carry out a similar linear stability analysis with the 
non-linear terms i.e. for the coupled circle maps as defined in Eq.2 with 
$K\neq 0$.
We consider a 1-d closed chain of $N$ lattice sites, with the sine circle map
at each site, coupled to its nearest neighbours via diffusive normalized 
symmetric
coupling and periodic boundary conditions, as in Eq.2, such that the right 
hand neighbour of the
$N$th point is the first lattice point.

 As discussed in the case of the coupled shift maps, for a
 synchronised solution, it makes sense to consider the evolution of the 
 differences. At any arbitrary time $t$ we write an equation
 similar to Eq.5, now for the coupled sine circle map. Consider, as an example,
 the first
 pair of lattice sites, of a lattice of N sites, with evolution at each site 
 defined as in Eq.2.
 \begin{eqnarray}
 \theta_{t+1}(1) - \theta_{t+1}(2) & = &
(1 - \epsilon)\Big\{\Big( \theta_{t}(1) - \theta_{t}(2)\Big) \nonumber \\
 & - & {K \over \pi}
 \sin \Big(\pi \Big(\theta_{t}(1) - \theta_{t}(2)\Big)\Big)
 \cos \Big(\pi \Big(\theta_{t}(1) + \theta_{t}(2)\Big)\Big) \Big\} \nonumber \\
 & + &{ \epsilon \over 2}\Big\{\Big( \theta_{t}(1) - \theta_{t}(N) \Big) + 
 \Big(
 \theta_{t}(2) - \theta_{t}(3)\Big) \nonumber \\
 & - &{K \over \pi}
 \sin \Big(\pi \Big(\theta_{t}(1) - \theta_{t}(N)\Big)\Big)
 \cos \Big(\pi \Big(\theta_{t}(1) + \theta_{t}(N)\Big)\Big) \nonumber \\
 & - &{K \over \pi}
 \sin \Big(\pi \Big(\theta_{t}(2) - \theta_{t}(3)\Big)\Big)
 \cos \Big(\pi \Big(\theta_{t}(2) + \theta_{t}(3)\Big)\Big)\Big\} 
 \hspace{0.04 in}mod \hspace{0.05 in}1 \nonumber \\
 \end{eqnarray}

Eq.12 shows, that for the coupled sine circle map, the evolution equation for 
differences
between the variable values of the nearest neighbours involves not just the 
differences between pairs of neighbours but also the sum of the 
variable values of the nearest neighbours,
 unlike the
coupled shift map where the evolution equation could be  defined in terms 
of the differences
alone. However, it is interesting to note that for a synchronised solution, 
 the difference of nearest neighbour lattice
sites is zero, and the sum of nearest neighbour lattice sites is a 
$constant$, 
for a fixed $\Omega$ and $K$. Therefore, we write a second
equation for the sums of the variables at nearest neighbour sites, 
and obtain,
 
 \begin{eqnarray}
 \theta_{t+1}(1) + \theta_{t+1}(2) & = &
 (1 - \epsilon)\Big\{\Big( \theta_{t}(1) + \theta_{t}(2)\Big) \nonumber \\
 & - & {K \over {\pi}}
  \sin \Big(\pi \Big(\theta_{t}(1) + \theta_{t}(2)\Big)\Big)
 \cos \Big(\pi \Big(\theta_{t}(1) - \theta_{t}(2)\Big)\Big) \Big\} \nonumber \\
   & + &{ \epsilon \over 2}\Big\{\Big( \theta_{t}(1) + \theta_{t}(N) \Big)
   + \Big( \theta_{t}(2) + \theta_{t}(3)\Big) \nonumber \\
     & - & {K \over {\pi}}
  \sin \Big(\pi \Big(\theta_{t}(1) + \theta_{t}(N)\Big)\Big)
   \cos \Big(\pi \Big(\theta_{t}(1) - \theta_{t}(N)\Big)\Big) \nonumber \\
& - & {K \over {\pi}}
\sin \Big(\pi \Big(\theta_{t}(2) + \theta_{t}(3)\Big)\Big)
\cos \Big(\pi \Big(\theta_{t}(2) - \theta_{t}(3)\Big)\Big)\Big\} +
\hspace{0.02 in} 2 \Omega \hspace{0.04 in} mod \hspace{0.01 in}1 \nonumber \\
  \end{eqnarray}
which again involves both the sums and the differences
of nearest neighbours. 
 It is clear that  for the synchronized solution
 of a lattice of coupled sine circle maps, we now need to define
 $N$ equations for the evolution of differences of nearest neighbour sites
 and $N$ equations for the evolution of sums of nearest neighbour sites as 
 compared to
 $N$ differences alone for the coupled shift map. This is due to the fact
 that the shift map is a linear version of the circle map and in the case
 of coupled circle maps 
 the non-linear sine term requires the identification of the second 
 independent variable.
 At any time $t$, we define the differences and sums, of the nearest
 neighbours, as follows,  
 \begin{equation}
 a_{t}(i)= \theta_{t}(i) - \theta_{t}(i+1) 
\end{equation}

 \begin{equation}
 b_{t}(i)= \theta_{t}(i) + \theta_{t}(i+1) 
 \end{equation}
\hspace{4.5 in} $ \forall \hspace{0.1 in}i ; 1, \ldots N $

Using Eq.2,  we find that
that $a_{t}(i)$ and $b_{t}(i)$ evolve via the following Eqs.  
\begin{eqnarray} 
a_{t+1}(i)& =& (1-\epsilon)\Big(a_{t}(i) - {K \over \pi} \sin( \pi a_{t}(i)) 
\cos( \pi b_{t}(i)i) \Big) \nonumber \\
&+& {\epsilon \over2}\Big(a_{t}(i+1) - {K \over \pi} \sin( \pi a_{t}(i+1)) 
\cos( \pi b_{t}(i+1)) \Big) \nonumber \\
&+& {\epsilon \over2}\big(a_{t}(i-1) - {K \over \pi} \sin( \pi a_{t}(i-1)) 
\cos( \pi b_{t}(i-1) \big) \nonumber \hspace{0.2 in} mod\hspace{0.05 in} 1\\
\end{eqnarray}
and
\begin{eqnarray} 
b_{t+1}(i)& =& (1-\epsilon)\Big(b_{t}(i) - {K \over \pi} \sin( \pi b_{t}(i)) 
\cos( \pi a_{t}(i)) \Big) \nonumber \\
&+& {\epsilon \over2}\Big(b_{t}(i+1) - {K \over \pi} \sin( \pi b_{t}(i+1)) 
\cos( \pi a_{t}(i+1))\Big)  \nonumber \\
&+& {\epsilon \over2}\Big(b_{t}(i-1) - {K \over \pi} \sin( \pi b_{t}(i-1)) 
\cos( \pi a_{t}(i-1)) \Big) \nonumber  + 2 \Omega  \hspace{0.1 in} 
mod\hspace{0.05 in} 1  \\
\end{eqnarray}	

For a synchronized solution, as mentioned above, at any time $t$, 

\begin{eqnarray}
a_{t}(i) & = & 0 \\
 b_{t}(i)& = & constant = s
\end{eqnarray}

  Inserting these conditions, Eq.16 reduces to
  \begin{equation}
       a_{t+1}(i) = 0
      \end{equation}

  implying that 
$  a_{t}(i) = 0 $ 
is a spatial fixed point solution.

 Similarly, Eq.17 reduces to
  \begin{eqnarray}
  b_{t+1}(i) = (1- \epsilon)(s-{K \over \pi} \sin( \pi s)) + \epsilon
  (s-{K \over \pi} \sin( \pi s)) + 2 \Omega \hspace{0.2 in} 
  mod\hspace{0.05 in} 1   
  \end{eqnarray}
  which is again a $ constant$ (not necessarily the same $constant$) 
  for a fixed $\Omega$ and $K$.

Eq.21 is  not restricted to  temporal evolution of any particular kind, 
implying that this same
equation may be used to study temporally periodic, quasiperiodic and 
chaotic behaviour. As an
example, $b_{t+1}(i) = $ the same $ constant \hspace{.01 in}s$ will 
indicate a temporal fixed point solution
and so on.

  Thus $a_{t}(i)$ = 0 and $b_{t}(i) = constant$ are
  solutions of the equations of evolution. 
 We now perform a Taylor expansion up to first order about these solutions
  to obtain the linear stability matrix $J_{t}$ which is of order 
  2$N \times2N$  and of the form
  
  \begin{equation}
   J_{t} = \pmatrix{ A ^\prime_{t} & B ^\prime_{t} \cr
   B ^\prime_{t} & A ^\prime_{t} \cr}
   \end{equation}

where 
\begin{equation}
 A^\prime_{t} = \pmatrix{ \epsilon_{s}A_{t}(1) &  
 \epsilon_{n} A_{t}(2) & 0 & 
  \cdots  & 0 & \epsilon_{n} A_{t}(N) \cr
  {\epsilon_n}A_{t}(1) &  
  {\epsilon_s}A_{t}(2) & 
  {\epsilon_n}A_{t}(3) & 0 & \cdots & 
  0 \cr
0 & 
\epsilon_{n}A_{t}(2) &
\epsilon_{s}A_{t}(3) &
   \cdots&  0 & 0 \cr
\vdots& \vdots& \vdots& \vdots & \vdots& \vdots  \cr       
  \epsilon_{n}A_{t}(1) & 0 &
   \cdots & 0 & 
  \epsilon_{n}A_{t}(N-1) &
  {\epsilon_s}A_{t}(N) \cr}
\end{equation}
and
\begin{equation}
  B ^\prime_{t} =    
\pmatrix{ \epsilon_{s}B_{t}(1) & 
  {\epsilon_n}B_{t}(2)  & 0 & \cdots &
   0 & {\epsilon_n}B_{t}(N)\cr 
 {\epsilon_n}B_{t}(1) &
{\epsilon_s}B_{t}(2) & 
 {\epsilon_n}B_{t}(3) & 0 &
  \cdots & 0 \cr
0 & {\epsilon_n}B_{t}(2) &
{\epsilon_s}B_{t}(3) & 
  \cdots &
  0 & 0 \cr
\vdots& \vdots& \vdots& \vdots& \vdots & \vdots  \cr       
 {\epsilon_n}B_{t}(1) & 0 & \cdots & 0 &
{\epsilon_n}B_{t}(N-1) & 
 {\epsilon_s}B_{t}(N) \cr} 
 \end{equation}

  Here
  \begin{equation}
 \epsilon_s = (1- \epsilon), \hspace{0.2in}
       \epsilon_n = {\epsilon \over 2}
       \end{equation}

 Each

\begin{equation}
       A_{t}(i)  =  {\Big (} 1 - {K \over \pi}\cos( \pi a_{t}(i))
      \cos( \pi b_{t}(i)) {\Big )} 
 \end{equation}
      and
      \begin{eqnarray}
       B_{t}(i)  =  {\Big (} 1- {K \over \pi}\sin( \pi a_{t}(i))
       \sin( \pi b_{t}(i)){\Big )}
\end{eqnarray}

Imposing the condition $ a_{t}(i)=0$, $ b_{t}(i) = s $
the stability matrix $J_{t}$, Eq.22 reduces to
\begin{equation}
    J_{t} = \pmatrix{ M_{t}& 0 \cr
       0& M_{t}\cr}
	  \end{equation}
	    where each $M_{t}$ is an $N \times N$ circulant matrix given by,
\begin{equation}
M_{t} = \pmatrix{ \epsilon_{s}\bar A_{t}(1) &  
 \epsilon_{n} \bar A_{t}(2) &  
  \cdots  & \epsilon_{n} \bar A_{t}(N) \cr
    \epsilon_{n}\bar A_{t}(1) &  
 \epsilon_{s} \bar A_{t}(2) &  
  \cdots  & 0  \cr
 0 &\epsilon_{n} \bar A_{t}(2) &  
  \cdots  & 0  \cr
\vdots& \vdots& \vdots& \vdots \cr       
    \epsilon_{n} \bar A_{t}(1) & 0 & \cdots& 
 \epsilon_{s} \bar A_{t}(N) \cr}   
\end{equation}
 For a synchronised solution all the lattice sites have the same value
 and hence each $b_{t}(i) = s $ $ \forall i, \hspace{0.02 in} 1, \ldots, N$. 
 Thus each
 \begin{equation}
  \bar A_{t}(i) = {\Big (} 1
   - {K \over \pi}
	 \cos( \pi s ) {\Big )}
 \end{equation}

   We thus have a block diagonal matrix
  with circulant blocks. In the stability analysis for other coupled map
  lattices carried out so far, 
  it was necessary to find a similarity
  transformation to obtain the stability matrix in block diagonal
  form \cite{gad 93}. In this case,
   we no longer have to find such a transformation, but directly 
  obtain the stability matrix in a block diagonal form. 

   To study the stability of the homogeneous solution, it is sufficient to 
   obtain the eigenvalues of one of the blocks $M_{t}$.

\vspace{0.1 in}

The eigenvalues of $M_{t}$ are given by Eq.9 \cite{gad 93,davis 79}, but are 
now of the form

\begin{eqnarray}
\lambda_{r} & = & (1- \epsilon)(1- K \cos( \pi s)) \nonumber \\
& + &
\omega_{r} {\Big(}{ \epsilon \over 2} (1- K \cos ( \pi s)){\Big)}
 + \omega_{r}^{N-1} {\Big(}{ \epsilon \over 2} (1- K \cos ( \pi s)){\Big)}
 \end{eqnarray}
    where $\omega_ {r}$ is defined as in Eq.10

	On simplifying, Eq.31 can be written as 
 \begin{eqnarray}
 \lambda_{r} = (1- \epsilon)(1- K \cos( \pi s))& +&
   \epsilon (1- K \cos ( \pi s)) \cos ({2 \pi (r - 1) \over N}), \nonumber\\
       && r = 1,2, \ldots, N
  \end{eqnarray}

Now, $\vert \lambda_{r} \vert < 1$, for all r, ensures the stability of the 
synchronized solution. The largest eigenvalue $ \lambda^{largest} $ 
crossing
1, is the condition for marginal stability \cite{gad 93}.

As mentioned earlier, we have
not assumed any temporal periodicity while defining $a_{t}(i)$ and
$b_{t}(i)$. They have been defined at any arbitrary time $t$. Thus, as
long as the condition for synchronization i.e. $a_{t}(i) = 0$ and
$b_{t}(i) = constant$ is satisfied, this formalism may be
used to identify and model any kind of synchronised temporal behavior. 

For the synchronized solution, we can analytically obtain those regions
in the $\Omega - K - \epsilon$ space, for which the spatially synchronized 
(i.e.
 spatial period 1), and additionally
temporal period 1 solution, is stable.

\subsection
{The Fixed Point case}
For the temporal fixed point, i.e. $a_{t+1}(i) = a_{t}(i) = 0$
and $b_{t+1} (i) = b_{t}(i) = s$, is
also a synchronised solution and hence also the spatial fixed point
of the system. Using Eq.32, the largest eigenvalue
of the stability matrix $J_{t}$ is given by
\begin{equation}
\lambda^{largest} = (1- K \cos (\pi s))
\end{equation}
Now, using the fixed point condition for Eq.(21),
\begin{equation}
s = s - {K \over \pi} \sin( \pi s) + 2 \Omega
\end{equation}

This gives
\begin{equation}
 s = {1 \over \pi} \arcsin ( {2 \pi \Omega \over K})
 \end{equation}
Inserting this value of $s$ and using the condition for
 stability, $\lambda \pm 1$, in Eq. 33, we obtain the width of the
 stability interval,

\begin{equation}
\Delta \Big( \Omega ({0 \over 1})\Big) = \Big({-K \over 2 \pi}, 
{K \over 2 \pi}\Big)
\end{equation}

This is the same width, as that obtained by Bak {\it et al}  \cite{bak 84} 
for the
single circle map. This width is the width of that region of 
$\Omega$, for which we obtain stable fixed point solutions. Thus, we find 
that
for the synchronized solution, the width of the $ 0 \over 1$ interval for
the coupled circle map case is identical to that obtained for
the single circle map \cite{bak 84}. However, in this case, it is the width 
of the of the
$0 \over 1$ interval for each site, and since the value of $\Omega$ at
each site is the same, for the entire lattice. Thus, for the region
of $\Omega$, defined by Eq.36, the entire lattice, has synchronized
stable temporal fixed point solutions.

For higher temporal periods, the stability matrix $J_{t}$, can be 
obtained in the following way. For a period $Q$ orbit, $J_{t}$ is simply
given by,
\begin{equation}
J_{t} = {\prod_{t = 1}^{Q}} \pmatrix{ A ^\prime_{t} & B ^\prime_{t} \cr
     B ^\prime_{t} & A ^\prime_{t}\cr}
\end{equation}    
where all the notation is the same as defined earlier.

Imposing, the condition for a synchronized solution, namely, Eqs.18
and 19,
$J_{t}$ reduces to the following form
\begin{equation}
J_{t} =  {\prod_{t = 1}^{Q}} \pmatrix{ M_{t} & 0\cr
     0 & M_{t}\cr}
\end{equation}	   
  The widths of the the higher temporal periods, for the synchronized case 
  cannot
be obtained analytically,as has previously been observed in the case of the 
single
circle map \cite{bak 84}. To calculate the widths of the higher temporal
periods, for the synchronized solution we carry out numerical simulations
which have been discussed in section IV. A detailed discussion is presented 
in
\cite{nc 95}.

 Since the
 formalism set up here is sufficiently general, it may also be used to 
 model
 synchronisation in other coupled map systems.

  The next section extends the formalism for solutions corresponding to 
  arbitrary spatial periods.
  
\section
{Analysis for the $k$th spatial period}
We find that the formalism discussed above may be efficiently used to
study the stability of higher spatial periods as well.
We extend it to higher spatial periods
 with the stability matrix retaining its block diagonal form.
  By higher spatial periods, say spatial period 4, we mean
   every $i$th and $(i+4)$th lattice site will have the same value.
    (See Fig.1b)

We now extend this formalism, to an arbitrary spatial period $k$. 
At any time $t$, for a spatial period $k$ solution, 
the value of the variable at the $i$th site is the same
as the value at the $(i + k)$th site. Thus, the difference, between the
variable values of the $i$th and the $(i + k)$th site, approaches zero, 
while the
sum of these variable values, approaches a constant.
As discussed in section II, now for a spatial period $k$, we can show 
that the differences and sums, are again the independent variables
of the system and are now defined as,
\begin{equation}
  a_{t}^{k}(i) = \theta_{t}(i) - \theta_{t}(i+k) 
 \end{equation}
\begin{equation}
b_{t}^{k}(i) = \theta_{t}(i) + \theta_{t}(i+k) 
 \end{equation}
where the superscript $k$ denotes the spatial period at any time $t$.

Eqs.16 and 17, now for a spatial period $k$, are modified to, 
\begin{eqnarray}
a_{t+1}^{k}(i)& =& (1-\epsilon)\Big(a_{t}^{k}(i) - {K \over \pi} 
\sin( \pi a_{t}^{k}(i)) \cos( \pi b_{t}^{k}(i)) \Big) \nonumber \\
&+& {\epsilon \over2}\Big(a_{t}^{k}(i+1) - {K \over \pi} 
\sin( \pi a_{t}^{k}(i+1)) \cos( \pi b_{t}^{k}(i+1)) \Big) \nonumber \\
&+& {\epsilon \over2}\Big(a_{t}^{k}(i-1) - {K \over \pi} 
\sin( \pi a_{t}^{k}(i-1)) \cos( \pi b_{t}^{k}(i-1)) \Big) \nonumber  
\hspace{0.2 in} mod\hspace{0.05 in} 1\\
\end{eqnarray}
and
\begin{eqnarray}
b_{t+1}^{k}(i)& =& (1-\epsilon)\Big(b_{t}(i) - {K \over \pi} 
\sin( \pi b_{t}^{k}(i)) \cos( \pi a_{t}^{k}(i)) \Big) \nonumber \\
&+& {\epsilon \over2}\Big(b_{t}^{k}(i+1) - {K \over \pi} 
\sin( \pi b_{t}^{k}(i+1)) \cos( \pi a_{t}^{k}(i+1)) \Big) \nonumber \\
&+& {\epsilon \over2}\Big(b_{t}^{k}(i-1) - {K \over \pi} 
\sin( \pi b_{t}^{k}(i-1)) \cos( \pi a_{t}^{k}(i-1)) \Big) \nonumber  + 
2 \Omega  \hspace{0.1 in} mod\hspace{0.05 in} 1\\
\end{eqnarray} 
       
 As shown in section II, here too it can be easily shown that, 
 $\forall {i},
 a_{t}^{k}(i) = 0$ and
    $ b_{t}^{k}(i) = s_{k} $, where $s_{1}, s_{2}, \ldots$ are all 
    $ constants$,
    and are solutions of the Eq. 41 and 42 for a fixed $\Omega$ and $K$.\\

 For a spatial period $k$ solution,
we expand about, $a_{t}^{k}(i) = 0$ and $ b_{t}^{k}(i) = constant $ and 
obtain the
stability matrix $J_{t}^{k}$\\
\begin{equation}
J^{k}_{t} =  \pmatrix{ A ^{\prime {k}}_{t} & B ^{\prime {k}}_{t}\cr 
   B ^{\prime {k}}_ {t} & A ^{\prime {k}}_{t}\cr}
   \end{equation}
where
\begin{equation}
 A ^{\prime {k}}_{t} = \pmatrix{ \epsilon_{s}A_{t}^{k}(1) &
 \epsilon_{n} A_{t}^{k}(2) & 0 &
 \cdots & 0 & \epsilon_{n} A_{t}^{k}(N) \cr
{\epsilon_n}A_{t}^{k}(1) &
 {\epsilon_s}A_{t}^{k}(2)  &
{\epsilon_n}A_{t}^{k}(3) &
    0  & \cdots & 0 \cr
    0 &
  \epsilon_{n}A_{t}^{k}(2) &
  \epsilon_{s}A_{t}^{k}(3) &
\cdots&  0 & 0 \cr 
 \vdots& \vdots& \vdots& \vdots& \vdots& \vdots  \cr
      \epsilon_{n}A_{t}^{k}(1) & 0 &
	 \cdots & 0 &
      \epsilon_{n}A_{t}^{k}(N-1) &
  {\epsilon_s}A_{t}^{k}(N) \cr}
 \end{equation}
 and
\begin{equation}
 B ^{\prime {k}}_{t} =  \pmatrix{ \epsilon_{s}B_{t}^{k}(1) &  
 \epsilon_n B_{t}^{k}(2) & 0 &
 \cdots &
 0 & {\epsilon_n}B_{t}^{k}(N) \cr
 {\epsilon_n}B_{t}^{k}(1) &
 {\epsilon_s}B_{t}^{k}(2) & 
 {\epsilon_n}B_{t}^{k}(3) & 0 &
   \cdots & 0 \cr
 0 & {\epsilon_n}B_{t}^{k}(2)
 & {\epsilon_s}B_{t}^{k}(3) 
  & \cdots &
      0 & 0 \cr
\vdots& \vdots& \vdots& \vdots& \vdots& \vdots \cr
  {\epsilon_n}B_{t}^{k}(1) & 0 & \cdots & 0 &
   {\epsilon_n}B_{t}^{k}(N-1) &
   {\epsilon_s}B_{t}^{k}(N) \cr}
\end{equation}
      where
     $\epsilon_s , \epsilon_{n}$ are defined by Eq.25
Here, each
 \begin{eqnarray}
 A_{t}(i) = {\Big (} 1 - {K \over \pi}\cos( \pi a_{t}^{k}(i))
  \cos( \pi b_{t}^{k}(i)) {\Big )}
  \end{eqnarray}
  and
 \begin{eqnarray}
 B_{t}(i) = {\Big (} 1- {K \over \pi}\sin( \pi a_{t}^{k}(i))
   \sin( \pi b_{t}^{k}(i)){\Big )}
\end{eqnarray}

Imposing the conditions $a_{t}^{k}(i) = 0$ and $ b_{t}^{k}(i) = s_{k} $, 
the stability matrix $J^{k}_{t}$ given by Eq.43 again reduces to 
a block diagonal form
\begin{equation}
J^{k}_{t} = \pmatrix{ M_{t}^{k}& 0 \cr
    0& M_{t}^{k}\cr}
	\end{equation}
	where each $M^{k}_{t}$ is of the form,
  \begin{equation}
 M^{k}_{t} = \pmatrix{ \epsilon_{s}\bar A_{t}^{k}(1) &
 \epsilon_{n} \bar A_{t}^{k}(2) &
 \cdots  & \epsilon_{n} \bar A_{t}^{k}(N) \cr
  \epsilon_{n}\bar A_{t}^{k}(1) &
 \epsilon_{n} \bar A_{t}^{k}(2) &
   \cdots \cr
0 &\epsilon_{n} \bar A_{t}^{k}(2) &
 \cdots  & 0  \cr
  \vdots& \vdots& \vdots& \vdots \cr
\epsilon_{n}\bar A_{t}^{k}(1) & 0 & \cdots&
 \epsilon_{s} \bar A_{t}^{k}(N)\cr} 
    \end{equation}
  and each
  \begin{equation}
    \bar  A^{k}_{t}(i) = {\Big (} 1
	- {K \over \pi}
		 \cos( \pi s_{m} ) {\Big )}
		 \end{equation}
where, each $ m $ goes from \hspace{0.1 in} $ 1, \ldots, k $ 
 \hspace{0.01 in}See Fig.1(b) for an illustration of spatial period 4.

Thus even for higher spatial periods, the stability matrix $J^{k}_{t}$ is
in  a block diagonal form, Eq.48.
Finding the eigen values of the matrix $  M_{t}^{k}$, and the largest eigen
value crossing + 1, gives the region of stability of the period
$k$ solution.

\section{\bf Numerical Simulations}
We obtained the width of $\Delta \Big( \Omega ({0 \over 1}) \Big) $, the 
temporal
period 1 case, for the
synchronized solution analytically in section II.
  The higher temporal periods,
   $\Delta \Omega ({P / Q}) $, for the synchronized solution, are obtained 
   numerically.

 The widths, $\Delta \Omega ({P / Q})$, as also stated in
 section II, are those regions of $\Omega$, at $K = 1$, for which, the
  entire lattice,
 will have temporally stable and spatially synchronized  periodic solutions of
time period $Q$.

 We extend the algorithm developed by Bak {\it et al} \cite {bak 84}, in the
  following manner.

 For a lattice of $N$ sites we define the following vector notation,
\begin{equation}
{\vec f}(\vec \theta, \vec \Omega) \rightarrow
\big\{ f_{1}(\vec \theta, \vec \Omega), f_{2}(\vec \theta, \vec \Omega), 
\ldots, f_{N}(\vec \theta, \vec \Omega)
     \big \}
 \end{equation}

   where
 $i$ denotes the lattice index and each $f_{i}(\vec \theta, \vec \Omega)$ 
 is defined by Eq.2
Each $\theta$ is now a vector
  of the form

\begin{equation}
 (\vec \theta) \rightarrow
  \big\{ \theta_{t}(1), \theta_{t}(2),  \ldots, \theta_{t}(N)  \big \}
  \end{equation}

 and the parameter, $\Omega$, also a vector, is represented as
\begin{equation}
 (\vec \Omega) \rightarrow
\big\{ \Omega (1),  \Omega (2), \ldots,  \Omega (N)  \big \}
 \rightarrow \big\{\Omega, \Omega, \ldots, \Omega \big\}
\end{equation}

 $\Omega$, in principle may
have different values at each site, but in this case has the same value at 
each site.
			  
For a 1-d array of coupled sine circle maps, i.e. a multi-dimensional case, 
the stability criterion is
obtained by examining
the eigen values of the following $ N \times N $ matrix
 \begin{equation}
 {S_t} =  {\prod_{t = 1}^{Q}} \pmatrix{ {\partial f_{1}} \over
  {\partial {\theta_{t}(1)}} &
 {\partial f_{1}}  \over
 {\partial { \theta_{t}(2)}} & \cdots & {\partial f_{1}}
 \over
 {\partial {\theta_{t}(N)}} \cr
 {\partial f_{2}} \over
{\partial {\theta_{t}(1)}} &
{\partial f_{2}}  \over
{\partial {\theta_{t}(2)}} & \cdots & {\partial f_{2}}
   \over
{\partial {\theta_{t}(N)}} \cr
\vdots & \vdots & \vdots & \vdots \cr
 {\partial f_{N}}  \over
 {\partial {\theta_{t}(1)}} &
 {\partial f_{N}}  \over
  {\partial {\theta_{t}(2)}} & \cdots & {\partial f_{N}}
   \over
  {\partial {\theta_{t}(N)}}\cr}
  \end{equation}
   Let $\{ \lambda_{i} \}$ be the set of eigenvalues of the matrix $S_{t}$.
   For a period $Q$ orbit to be stable the eigenvalues of the
    matrix $S_{t} < 1$.
 The largest eigenvalue crossing 1 defines
 the marginal stability condition.
  
  Thus for the higher order temporal periods,
  we seek the solution
   to the set of Eqs. 55 and 56,
 \begin{equation}
  {\vec f^Q}(\vec \theta, \vec \Omega)  =   (\vec \theta) + {\vec P}
   \end{equation}
  the condition for closure, and
		   
   \begin{equation}
   \lambda^{largest}  =   1
   \end{equation}
    the condition for marginal stability
		     
  For a general $P / Q$ step, and a lattice of $N$ sites, we start with
  homogeneous initial conditions and perform the
iterative technique based on the Newton - Raphson method.
					 
 We define
   \begin{eqnarray}
\vec g_{1}( \vec \theta, \vec \Omega ) & = & \{ g_{1}(1), g_{1}(2), \ldots, g_{1}(N) \} \\
 \vec g_{2}( \vec \theta, \vec \Omega )  & = & \{ {{ \partial { f_{1}^{Q}}}
 \over {\partial {\theta_{t}(1)}}}, {{\partial {f_{2}^{Q}}} \over {\partial {\theta_{t}(2)}}}, \ldots, {{\partial { f_{N}^{Q}}} \over {\partial {\theta_{t}(N)}}} \}
\end{eqnarray}
where  
\[ \vec g_{1}(1) = {f_{1}^Q}(\vec \theta, \vec \Omega)
 - (\theta(1)) - {\vec P} \] and so on.
						 
 We further define,

 \begin{equation}
   {\vec g} ( \vec \theta, \vec \Omega ) = { \vec {g_1} \Big( \vec \theta, 
   \vec  \Omega  \Big) \choose
  \vec {g_2}\Big(  \vec\theta,  \vec \Omega  \Big)}
   \end{equation}
 where now,
 $$\vec g_{2}( \vec \theta, \vec \Omega )   =  {\vec g_{2}( \vec \theta, 
 \vec
 \Omega )} - \vec 1 $$
	  
 For the synchronized solution, it can be shown, that the largest eigenvalue 
 of the matrix, $S_{t}$, namely $ \lambda^{largest}$, is equal to each of the 
 $N$ components of $\vec g_{2}$ in Eq.58 \cite{nc 95}.
		    
 Therefore, Eqs.55 and 56,
 may now be expressed as
 finding $( \vec \theta^{\ast}, \vec \Omega^{\ast} )$ such that
 \begin{eqnarray}
 {\vec g}{(\vec \theta^ \ast, \vec \Omega^\ast)}  = \vec g^{\ast} = \vec 0
 \end{eqnarray}
			       
Taylor expanding $ \vec g^{\ast}$, about the initial point of iteration, 
$\vec g ( \vec \theta^{0},
 \vec \Omega^{0} ) = \vec g_{0} $, up to the
  linear order, we obtain,
  \begin{equation}
 \vec{g}{\ast}  \simeq {\vec {g_0}}
 + {\vec {\Delta}}{ M}
 \end{equation}
 where
 \begin{equation}
 {\vec {\Delta}} =  \Big(\vec \theta^{\ast}, \vec \Omega^{\ast} \Big)
 -\Big(
  \vec \theta^{0}, \vec \Omega^{0} \Big)
   \end{equation}

\begin{equation}
  M = \pmatrix{ {\partial { {g_1}(1)}} \over {\partial \theta_{t}(1)} &  
  \cdots & {\partial { {g_1}} (1) }  \over {\partial \theta_{t}(N)} & 
  {\partial { {g_1}} (1) }  \over {\partial \Omega (1) } &  \cdots & 
  {\partial { {g_1}} (1) } \over {\partial \Omega (N) } \cr
   \vdots & \vdots & \vdots & \vdots & \vdots & \vdots \cr
    {\partial { {{g_1}}}(N)} \over {\partial \theta_{t}(1)} &  \cdots &
    {\partial { {g_1}}(N)}  \over {\partial \theta_{t}(N)} & {\partial 
{ {g_1}} (N)} \over {\partial \Omega (1) } &  \cdots & {\partial { {g_1}} (N)} 
\over { \partial
 \Omega (N) } \cr
{\partial {{g_2}}(1)} \over {\partial \theta_{t}(1)} & \cdots & 
{\partial {{g_2}}(1)}  \over {\partial \theta_{t}(N)} & {\partial {{g_2}} (1)} 
\over
{\partial \Omega (1) } &  \cdots & {\partial {{g_2}} (1)} \over 
{\partial \Omega (N) }\cr
\vdots & \vdots & \vdots & \vdots & \vdots & \vdots \cr
 {\partial {{g_2}}(N)} \over {\partial \theta_{t}(1)} & \cdots & 
 {\partial {{g_2}}(N)}  \over {\partial \theta_{t}(N)} & {\partial {{g_2}} (N)} \over
{\partial \Omega (1)} &   \cdots & {\partial {{g_2}} (N)} \over 
{\partial \Omega (N)} \cr}
 \end{equation}
It is found that, for
 \begin{equation}
   {\vec{g}^{\ast}} = {\vec{0}}
\end{equation}
\begin{equation}
 {\vec {\Delta}} \simeq -{ {M}}^{-1}{\vec g_{0}}
  \end{equation}
And so as a first approximation,
\begin{equation}
{ \vec \theta^\ast ,\choose \vec  \Omega^\ast}  \simeq {\vec \theta^1  
\choose  \vec \Omega^1}
  =-{{M}}^{-1}{\vec g_{0}} + { \vec \theta^0 \choose \vec \Omega^0}
  \end{equation}
		      
  All derivatives can be derived recursively \cite{nc 95}

We start with homogeneous initial conditions, for the synchronized case, 
and find, to initiate the iteration, it is convenient
to locate the superstable point $(\vec \theta, \vec \Omega) = 
(\vec \theta_s,
\vec \Omega_s)$, for all the lattice sites, where now the eigenvalues of the 
matrix $S_{t}$ are = 0.
All steps have been
found to an accuracy of $10^{-8}$.
	 
Using the alogrithm discussed above, at $K=1$ we obtain the complete 
devil's Staircase, of periodic orbits,
 now
 in the $ \Omega - {P \over Q} - \epsilon $
 space (see Fig.2) with all the special features seen in the single circle map.
	  
  We also studied various features of the coupled
  sine circle map lattice at lower values of $K$ and different $\Omega$'s. 
  Our simulations reveal synchronized quasi-periodic orbits and periodic 
  orbits in
  the $\Omega - K- \epsilon $ space.We thus obtain the entire structure
  of the Arnold tongues ( see Fig.3) now with the third coupling dimension 
  $\epsilon$. It is clear from ( see Fig. 3 that the three
possible routes to chaos seen in the case of the single
circle map i.e. via mode-locking alone , via quasi-periodicity and 
mode-locking and via quasi-periodicity to chaos, will also be seen for the 
coupled sine circle map lattice.
	   
   Thus for the synchronized solution, we obtain all the features of the
   single map, now with a third dimension, in the form of the
   coupling parameter. We also observe that for the synchronised solution all 
   the features as seen in the
   single circle map, are lifted
   exactly to this extra dimension in the parameter space, namely the coupling
   parameter $\epsilon$, for the entire range,
    $ 0 \leq \epsilon \leq 1$.

  \section {CONCLUSIONS} We have set up a formalism to analyze
  the phenomenon of synchronization in a lattice of coupled sine circle maps. 
 We identify the independent variables, the differences and the sums
 of variable values at neighbouring sites, $a_{t}(i)$ and $b_{t}(i)$ for the 
 problem and show that, $ a_{t}(i) = 0$ and $b_{t}(i) = constant$,
 the synchronised solution, are solutions, of the system under consideration. 
 Casting the
 evolution equations in terms of these independent variables 
 leads to a stability matrix which is in a neat block
 diagonal form. In addition, these blocks have a circulant structure,
 so we have a block diagonal matrix with circulant blocks. The
 temporal fixed point case, for the synchronised solution, i.e. the spatial 
 period 
 one and temporal period one solution can be solved explicitly to obtain the 
 corresponding width
 of the stability interval. We have also demonstrated that this analysis can 
 be easily extended
 to higher spatial periods. For an arbitrary spatial period $k$, we identify 
 the
 independent variables, now $a^{k}_{t}(i)$ and $b^{k}_{t}(i)$ and again show 
 that
  $a^{k}_{t}(i) = 0$ and $b^{k}_{t}(i) = k \hspace{0.01 in}constants$, are 
  solutions of the system. A linear stability analysis for this case shows 
  that the stability matrix retains its block
 diagonal form. This analysis may also be easily extended to spatially 
 periodic
 behaviour in other coupled
 map systems. The analysis is one of the simplest presented so far with the 
 additional advantage of being transparent and neat.
 
 We calculate the widths of the higher temporal periods for the synchronized 
 case,
  by extending the algorithm developed by Bak{\it et al} \cite{bak 84}. 
  Our numerical simulations reveal that the for the synchronised case, 
  the entire structure of the
 Arnold Tongues and the devil's Staircase as observed for the single circle 
 map is lifted to the chain of coupled sine circle maps. 
 
A study of the parameter basins of attraction, i.e. those regions in
the $ \Omega - \epsilon $, where a particular class of initial conditions
converges to the synchronized solution, which is the attractor, have also
been carried out. These results which have been presented elsewhere 
 \cite {nc 95} showed that
these 
regions were found to be fairly large and showed an interesting
symmetry about $\Omega = 0.5 $. Random initial conditions also
revealed a finite basin of attraction, which was also symmetric
about $\Omega = 0.5 $. The numerical
simulations for the random initial conditions showed that
there always exists a finite region of the 
$ \Omega - \epsilon$ parameter space, at $K = 1$, for which synchronized
solutions will be obtained.
 The temporal period for these synchronized
regions was found to be 1.\cite{nc 95}.

 Investigation of higher spatial periods reveals extremely interesting
 structures, results of which will be presented elsewhere. 
Our analysis was carried out for the homogenous case which assumed  that the parameter values at each 
lattice site were  the same. However the formalism set up is completely general and can be easily extended to inhomogeneous cases. This should be useful for the study of lattices
where some lattice sites function as pinning sites. 
 
 We hope this work will find useful applications in the study of spatially
 periodic behaviour in 
 systems  like  
 Josephson junction arrays and coupled oscillator arrays and in studies of pattern formation, in reaction-diffusion systems and charge-density
 waves. 

 \vspace{0.2 in}
 
 {\bf ACKNOWLEDGMENTS}

\vspace{0.2 in}

 One of the authors (NG) acknowledges UGC and DST
  India, and (NC) gratefully acknowledges CSIR (India), for financial support.

\newpage
\begin{center}

\end{center}
\newpage

\begin{center}
{\bf FIGURE CAPTIONS}
\end{center}
\vspace{1.0 in}
\begin{center}
\begin{tabular}{lll}
 FIG.1 & (a) & The synchronised or spatial period 1 solution\\
 & (b) &  The spatial period 4 solution\\
 && Each symbol in the figure represents a distinct value of the
 variable of a lattice of 8 sites at any time $t$.\\

&&\\

 FIG.2 &&  The Devil Staircase for coupled sine circle maps.\\
 &&  The symmetry about
 $\Omega = 0.5$ is clearly seen.\\ &&The widths are calculated for 
 $\Delta \Omega ({P \over Q})$ for\\ && ${ P \over Q}$ = $0 \over 1$,
 ${1 \over 1}$,${ 1 \over 2}$, ${1 \over 3}$, ${2 \over 3}$,
 ${1 \over 4}$, ${3 \over 4}$, ${1 \over 5}$, ${2 \over 5}$.\\
  && This plot is for 8 lattice sites, but
 for a synchronized solution,\\ && identical 
 results are seen for any $N$.\\

&&\\
      
 FIG.3 && Plot of the Arnold  tongues, for the coupled sine
circle map lattice. \\&& The tongues have been plotted
for $\Omega = {P \over Q}$ starting from \\ &&  ${ P \over Q}$ = $0 \over 1$,
 ${1 \over 1}$,${ 1 \over 2}$, ${1 \over 3}$, ${2 \over 3}$\\ 
 && This plot is for 8 lattice sites, but
 identical results are seen for any $N$, \\ && for the synchronized
  solution.\\
\end{tabular}

 \end{center}

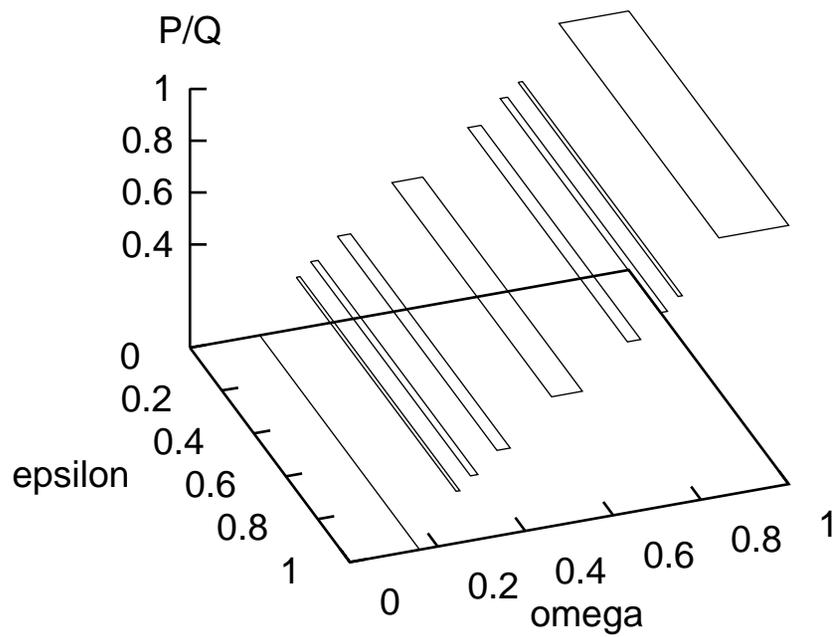
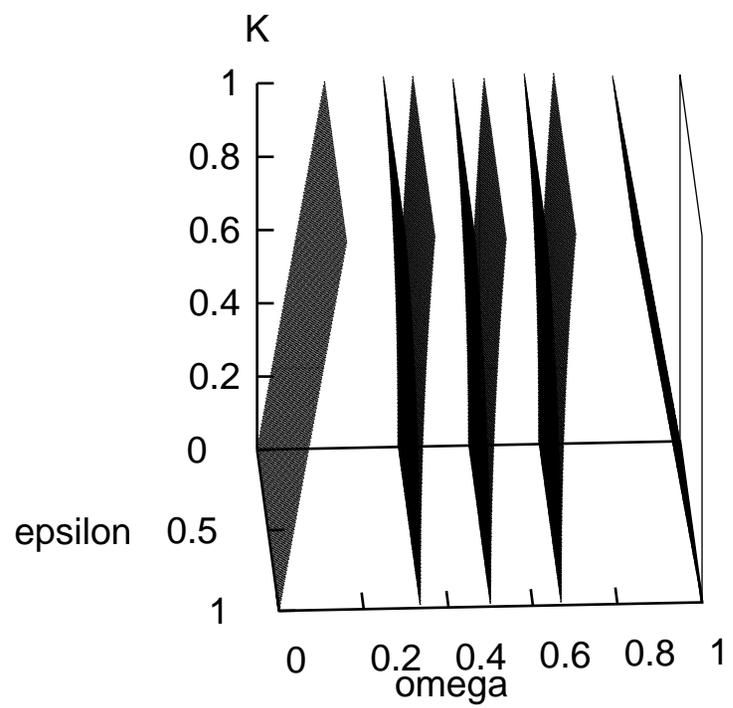
\end{document}